\documentclass[conference]{IEEEtran}
\usepackage{blindtext}
\usepackage{graphicx}
\usepackage{adjustbox}
\usepackage{algorithm}
\usepackage{algpseudocode}
\usepackage{algorithmicx}
\usepackage{amsmath}
\usepackage{epsfig}
\usepackage{booktabs}            % tables

\usepackage{url} %url reference
\usepackage{etoolbox}
\apptocmd{\thebibliography}{\raggedright}{}{} %url過長自動換行

\begin{document}
\title{Enhancing Modbus TCP Protocol Security with eBPF Technology}

\author{
\IEEEauthorblockN{Jia-Yi Jhan}
\IEEEauthorblockA{National Tsing Hua University\\
Email: s111138501@m111.nthu.edu.tw}\\
\and
\IEEEauthorblockN{Hung-Min Sun}
\IEEEauthorblockA{National Tsing Hua University\\
Email: hmsun@cs.nthu.edu.tw}
}

\maketitle

\begin{abstract}
%\boldmath
The core component of an Industrial Control System (ICS) is often a Programmable Logic Controller (PLC) combined with various modules. In such systems, the communication between devices is mainly based on the Modbus protocol, which was developed by Modicon (now Schneider Electric) in 1979 as an application-level communication protocol and has become \textit{a de facto standard} for ICS for the past 40 years. Modbus TCP is a variant of this protocol for communications over the TCP/IP network. However, the Modbus protocol was not designed with security in mind, and the use of plaintext transmissions during communication makes information easily accessible to the attackers, while the lack of an authentication mechanism gives any protocol-compliant device the ability to take over control.

In this study, we use the eBPF technology to shift the process of protocol change to the lower level of the operating system, making the change transparent to the existing software, and enhancing the security of the Modbus TCP protocol without affecting the existing software ecosystem as much as possible.
\end{abstract}

\begin{IEEEkeywords}
eBPF, Modbus, Industrial Control System
\end{IEEEkeywords}

% Introduction ===================================================================================
\section{Introduction}
\subsection{Motivation}
The Industrial Control System (ICS) serves as a specialized computer system tailored for the monitoring and management of industrial processes, with its primary function being the surveillance, control, and automation of various operations within factories, facilities, and industrial processes across sectors like manufacturing, energy, and the chemical industry. A key component within the Industrial Control System is the Programmable Logic Controller (PLC), employed to oversee and regulate both mechanical and electronic equipment throughout the manufacturing process. Its central role involves the automatic control of machine operations or processes based on predefined logic programs and input signals. The primary communication method employed by industrial control equipment is the Modbus communication protocol. Developed by Modicon Corporation (now Schneider Electric) in 1979, Modbus stands as an application layer communication protocol and has evolved into the industry standard for Industrial Control Systems over the past four decades. Modbus TCP represents a variant implementation of this protocol on the TCP/IP network.

In the past, these critical devices operated in isolation, confined to separate networks without direct Internet connectivity. With the ongoing wave of digital transformation, there is an observable trend towards integrating more industrial control devices with the Internet. Nevertheless, this shift introduces heightened risks. Modbus TCP, a longstanding communication protocol, has been widely adopted, yet its security considerations have garnered increasing attention. In contrast to traditional security measures, which may prove insufficient in the face of contemporary malicious attacks, especially in the absence of robust encryption and authentication mechanisms, Industrial Control Systems find themselves potentially more vulnerable. In light of this, there emerges a pressing need to address and enhance the security aspects of the Modbus TCP protocol.

Prior research endeavors have aimed to enhance the security of Modbus TCP by altering the protocol format. However, these approaches often necessitate extensive modifications to existing software and infrastructure, posing challenges for practical implementation within real-world environments.

This research is motivated by a desire to offer a comparatively practical and feasible methodology, shifting the emphasis of security enhancement to the “lower” layer of the operating system through the utilization of eBPF technology. This approach not only serves to fortify Industrial Control Systems against emerging threats but also facilitates improvements without causing disruption to the established software ecosystem.

\subsection{Contribution}
The proposed approach in the thesis may present a relatively pragmatic method for enhancing the security of the Modbus TCP protocol. While numerous research studies have suggested altering the protocol format to improve Modbus TCP security, introducing new protocols into the established infrastructure and software of Industrial Control Systems poses significant challenges. For instance, refactoring the entire software framework of a power station simply to implement a new Modbus protocol is not a feasible solution. In contrast, applying an eBPF program to devices, although a challenging task, is comparatively more manageable and feasible than previous alternatives. In practical implementation, we deploy the eBPF program on both the egress and ingress paths of the device. When a plaintext Modbus TCP packet is transmitted, the eBPF program encrypts the packet, and conversely, it decrypts an encrypted Modbus TCP packet upon reception. Importantly, the encryption and the decryption process remain transparent to the software. That is, existing software can be leveraged without extensive refactoring.

\subsection{Organization}
The proposal is organized as follows. In Chapter 2 we introduce the basics of the Modbus TCP protocol, a brief introduction to eBPF, and the two subsystems based on eBPF: XDP and TC. In Chapter 3 we present the existing methods to enhance Modbus security.

% Background ===================================================================================
\section{Background}
\label{back}

\subsection{Modbus TCP Frame Structure}
\begin{figure}[h]
    \centering
    \includegraphics[width=1\linewidth]{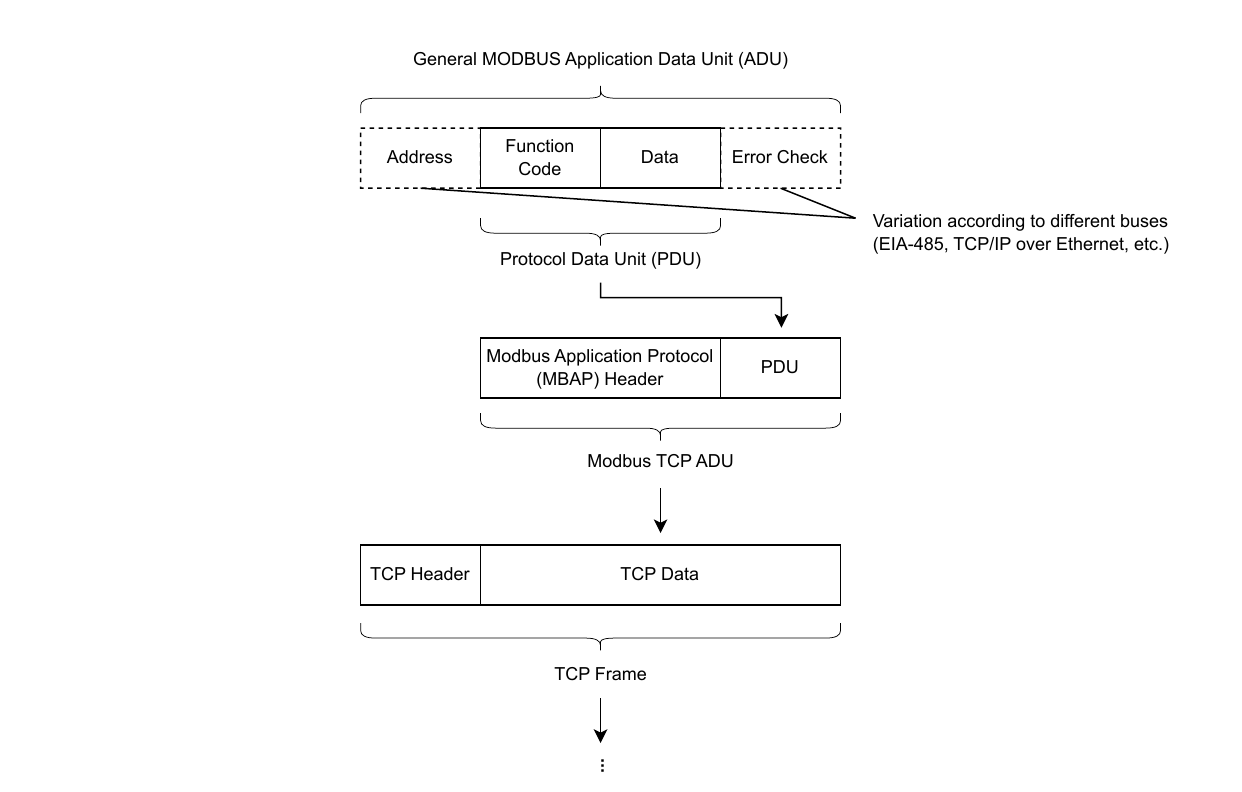}
    \caption{Modbus TCP Frame}
    \label{fig:modbus_tcp_frame}
\end{figure}

Modbus TCP frame structure is shown as Figure \ref{fig:modbus_tcp_frame}. Some fields would be discarded or alternated depending on what bus the protocol is running on. In detail, the general Modbus frame defines two data units: protocol data unit(PDU) and application data unit(ADU). The PDU consists of a function code field and a data field, which is used to interchange the message between the client and server. The ADU contains the corresponding header and tail information of the PDU (Protocol Data Unit) and the underlying transmission protocol used for communication. In Modbus TCP, the address and error check field are replaced with the Modbus Application Protocol(MBAP) header and Ethernet TCP/IP link layer checksum respectively.
%% Modbus TCP Frame

\subsection{Modbus TCP Function Code}
The Modbus TCP function code field is encoded in a single byte, allowing for codes within the valid range of 1 to 255. The range of 128 to 255 is reserved and used for exception responses. These function codes can be categorized into three types: Public Function, User-Defined Function, and Reserved Function.

Public Function Codes within the Modbus TCP protocol are well-defined and ensure uniqueness. In contrast, User-Defined Function Codes are utilized to implement functions not supported by the standard specification. However, employing User-Defined Function Codes may raise compatibility concerns, given the absence of a guarantee that the selected function code will remain unique across various implementations. Reserved Function Codes are not intended for public use since they are utilized to support legacy devices within specific vendors. \cite{Modbus-Application-Protocol-Specification}
%% Modbus Function Code Categories
\begin{table}[h]
\centering
\caption{Modbus Function Code Categories}
\begin{tabular}{@{}cl@{}}
\toprule
Function Codes & \multicolumn{1}{c}{Categories}                    \\ \midrule
1 - 64         & Public Function                                   \\
65 - 72        & User-Defined Function                             \\
73 - 99        & Public Function                                   \\
100 - 110      & User-Defined Function                             \\
111 - 127      & Public Function                                   \\
128 - 255      & Exception Responses (Request function code + 128) \\ \bottomrule
\end{tabular}
\end{table}

\subsection{eBPF}
The extended Berkeley Packet Filter(eBPF) can considered a sandbox running within the kernel, making users dynamically enhance the operating system's functionalities by running the eBPF programs. This runtime extensibility allows for the addition of supplementary capabilities to the operating system securely and efficiently. The operating system ensures safety and execution efficiency by leveraging a Just-In-Time (JIT) compiler and verification engine, treating eBPF programs as if they were natively compiled. This flexibility has given rise to a proliferation of eBPF-based projects spanning a diverse range of applications, including next-generation networking, observability tools, and security functionalities, showcasing the versatility and utility of eBPF across various use cases.

The eBPF leverages event-driven programming, creating multiple hooks that execute when specific hook points are triggered. Users can create either a kernel probe (kprobe) or a user probe (uprobe) to hook their eBPF programs at specific points within the kernel or user programs. The framework provides a set of predefined hooks tailored to handle common scenarios, including hooks for system calls, network events, function entry/exit points, and kernel tracepoints.
%% eBPF hook overview
\begin{figure}[h]
    \centering
    \includegraphics[width=1\linewidth]{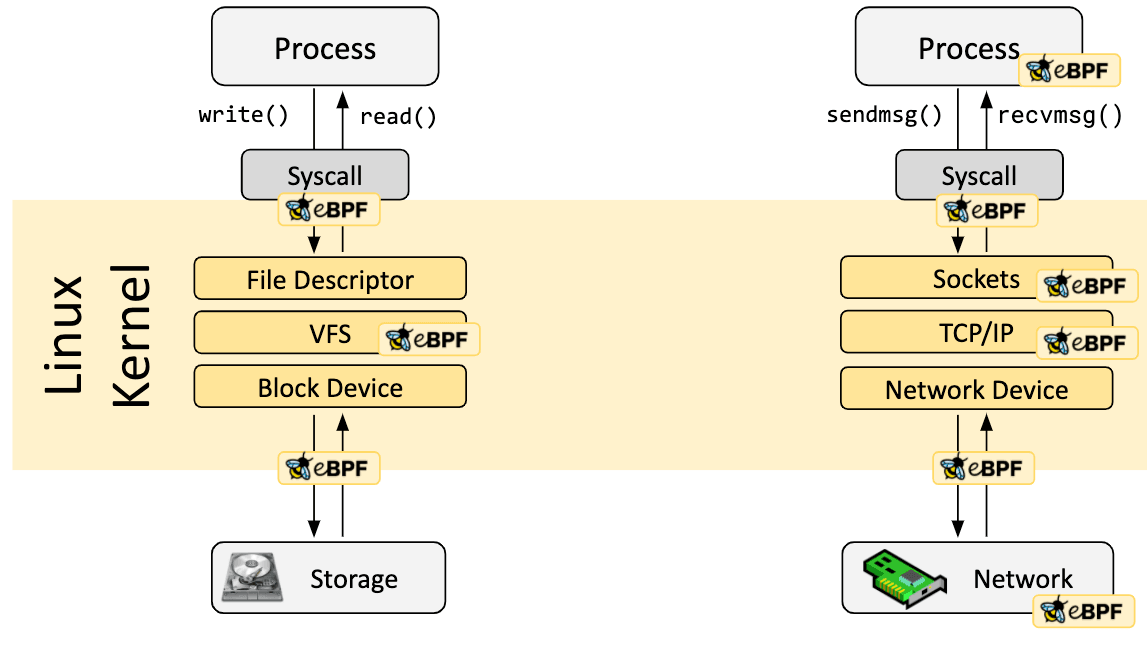}
    \caption{eBPF hook overview. \cite{eBPF-Documentation}}
    \label{fig:ebpf-hook-overview}
\end{figure}

To ensure the secure and efficient execution of eBPF programs, the kernel employs two methodologies: the eBPF verifier and the in-kernel Just-In-Time (JIT) compiler. The eBPF verifier conducts a meticulous static analysis of the eBPF program bytecode, aiming to establish its safety. Specifically, the verifier checks for the absence of:
\begin{enumerate}
  \item Infinite loops.
  \item Use of uninitialized variables or access to invalid memory.
  \item Exceed the program size requirements of the system.
  \item Unpredictable execution paths.
\end{enumerate}
While the eBPF program passes the check, it can be loaded safely. The kernel calls the \textit{bpf\_prog\_select\_runtime} function to specify the runtime for the eBPF program. On mainstream architectures like X86 or ARM64, the preferred runtime is typically JIT compilation, compile eBPF bytecode into target architecture machine instructions to improve performance. In scenarios where the architecture is not supported, or when users opt to disable the eBPF JIT function, the eBPF bytecode interpreter comes into play.
%% eBPF architecture
\begin{figure}[h]
    \centering
    \includegraphics[width=1\linewidth]{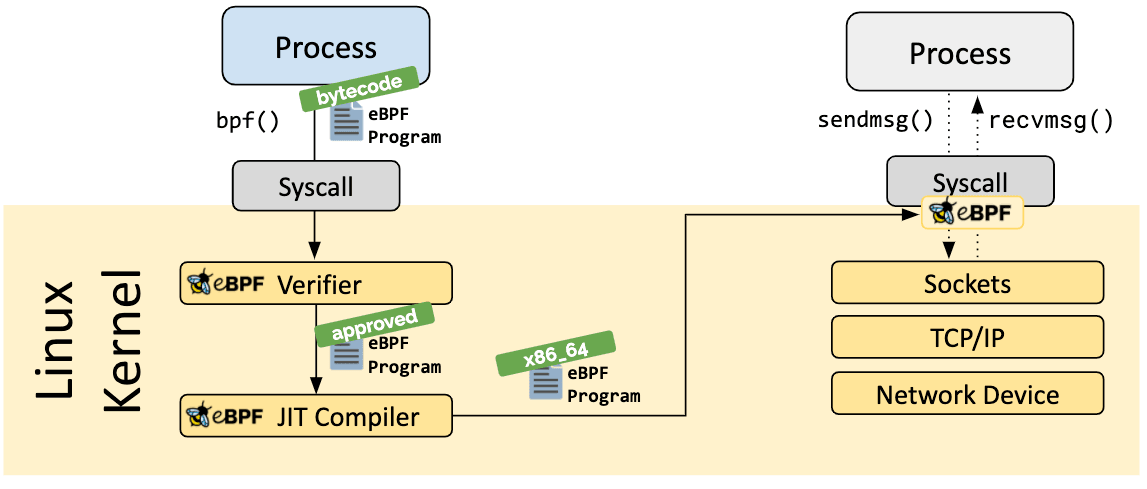}
    \caption{eBPF architecture. \cite{eBPF-Documentation}}
    \label{fig:ebpf-loader}
\end{figure}

\subsection{eBPF Maps}
Each execution of an eBPF program initiates from a consistent initial state. The concept of eBPF maps is introduced to empower eBPF programs with the capability to efficiently store and retrieve data. These eBPF maps function as key/value stores and are instantiated during the loading of eBPF programs. Additionally, the maps can be pinned to the eBPF filesystem during creation, allowing users to access them through file descriptors. Furthermore, different programs can parallelly access the same maps. This flexibility extends the use of maps beyond the initial purpose of data storage, making them applicable for data exchange between user space and kernel space, as well as facilitating shared data among multiple programs.
%% eBPF map architecture
\begin{figure}[ht]
    \centering
    \includegraphics[width=1\linewidth]{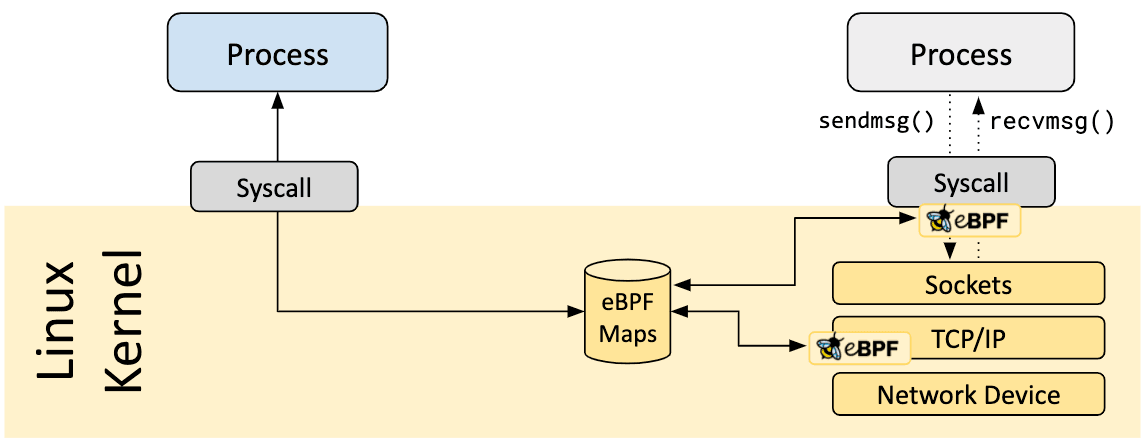}
    \caption{eBPF maps architecture. \cite{eBPF-Documentation}}
    \label{fig:ebpf-map}
\end{figure}

\subsection{eXpress Data Path}
The eXpress Data Path(XDP) is a programmable packet processing framework that is based on eBPF. In traditional packet processing, operations are typically performed after creating a \textit{sk\_buff} structure in the kernel, as seen in applications such as HAProxy \cite{HAProxy}, a user space application operating at OSI Layer 7, and the IP Virtual Server (IPVS) \cite{IPVS}, a Netfilter module functioning at OSI layer 4. In contrast, XDP programs integrate with the network driver, allowing direct access to raw packets from the receive ring buffer for packet processing. This eliminates the need to wait for the kernel to create a \textit{sk\_buff} structure and pass it to the network stack for processing, resulting in optimized performance utilization.
%% XDP structure
\begin{figure}[h]
    \centering
    \includegraphics[width=0.65\linewidth]{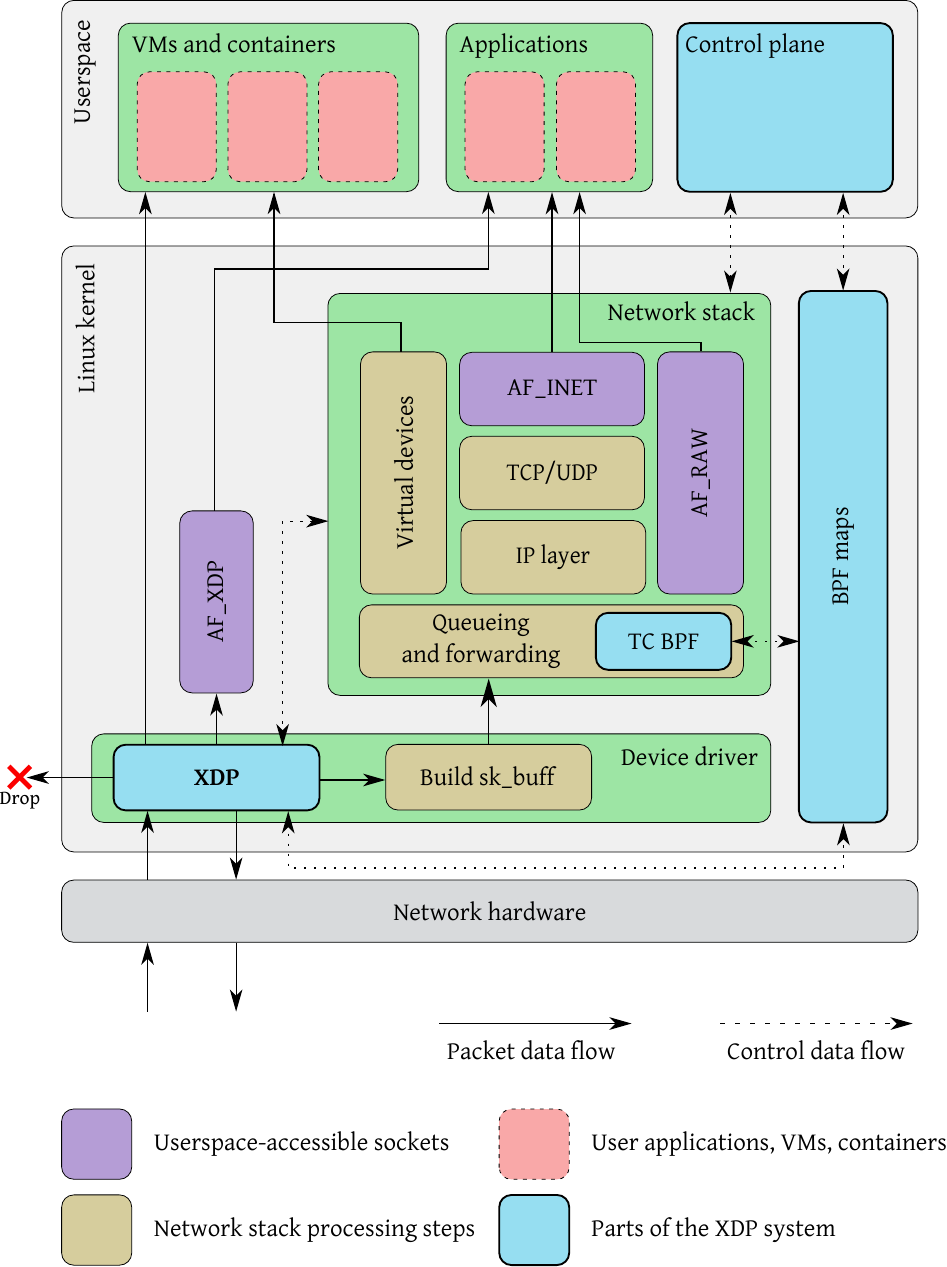}
    \caption{XDP's intergration with the Linux network stack. \cite{The-EXpress-Data-Path}}
    \label{fig:kernel-diagram}
\end{figure}

XDP Action determines the behavior when a packet is received, there are five actions:
\begin{enumerate}
    \item XDP\_PASS: Forward the packet to the kernel stack.
    \item XDP\_DROP: Drop the packet without raising an error. This is the fastest way to drop the packet.
    \item XDP\_ABORT: Drop the packet while raising an error. The distinction between XDP\_DROP and XDP\_ABORT lies in their usage. Typically, XDP\_DROP is used to indicate that a packet is in a valid format and can be correctly parsed, but it needs to be discarded due to triggering protection rules. On the other hand, XDP\_ABORT signifies an error during the execution of the eBPF program, resulting in the inability to parse the packet correctly, necessitating its discard.
    \item XDP\_REDIRECT: Pass the packet to the specific CPU, NIC, or User space socket(Kernel bypass).
    \item XDP\_TX: Pass the packet to same(i.e., ingress) NIC. Usually, the packet is modified when it is returned.
\end{enumerate}

\subsection{Traffic Control}
To effectively manage and shape network traffic, the Traffic Control (TC) subsystem was introduced, comprising two essential components: the TC framework within the Linux kernel and the user space interface for configuring the framework. In kernel 4.1, new hooks for eBPF programs were added, allowing eBPF programs to run as TC filters or TC actions. In kernel 4.4, iproute2\cite{iproute2} introduced a new flag known as direct-action (da), which makes the TC subsystem treat the return value of TC filters as a TC action. In kernel 4.5, iproute2 introduced the new type of qdisc called \textit{clsact}, bringing the TC direct-action feature to egress qdisc. We will provide further elucidation on these terms and components in the following section. There are four of the most important objects in the TC subsystem:
\begin{enumerate}
    \item Queuing Discipline: Abbreviated as qdisc. Applying different policies to “shape” the network traffic. It uses the FIFO algorithm to schedule by default.
    \item Class: Categorize the qdisc according to the different responsibilities(User-defined).
    \item Filter: Dispatch the packet to the corresponding classes. The return value of a filter can be the following one:
        \begin{itemize}
            \item 0: Mismatch, the packet will checked by the next filter.
            \item -1: Pass to the default class of the filter.
            \item Other value \textit{n}: Pass to the class specified by the classid \textit{n}.
        \end{itemize}
    \item Action: Optional. Defining the behavior independently of a filter, such as dropping the packet directly.
\end{enumerate}

\begin{figure}[ht]
    \centering
    \includegraphics[width=1\linewidth]{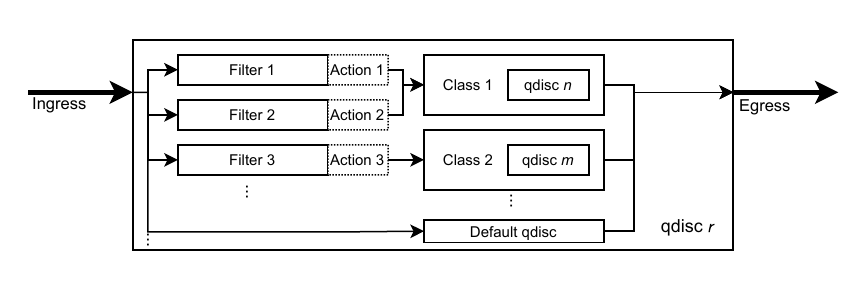}
    \caption{TC qdisc architecture.}
    \label{fig:tc-qdisc-architecture}
\end{figure}

The architecture of TC qdisc is illustrated in Figure \ref{fig:tc-qdisc-architecture}. We can notice that a qdisc can contain more than one class, and reciprocally, a class can contain a qdisc. It makes the whole qdisc structure consist of a tree structure, with the root as the qdisc \textit{r}.
%% TC qdisc architecture

The TC subsystem provides two hooks for eBPF programs: TC filter and TC action. The hooks make eBPF programs be used as a filter or action. While the eBPF program is used as a TC filter, it returns the same value as a classical filter and works as the normal filter but provides more flexibility such as using the eBPF map to store the data log, or dispatch to the different classes according to the data in the map. For the case of using the eBPF program as a TC action, the eBPF program return value decides what action should be performed on the packet. Following are some common return values \cite{TC-eBPF}:
\begin{enumerate}
    \item TC\_ACT\_UNSPEC(-1): Use default action from classical TC subsystem.
    \item TC\_ACT\_OK(0): The processing chain is ended and the packet is accepted.
    \item TC\_ACT\_RECLASSIFY(1): Start over, and re-execute the classification process from the beginning.
    \item TC\_ACT\_SHOT(2): Drops the packet and ends the processing chain.
    \item TC\_ACT\_PIPE(3): The packet is passed to the next action. If there is a next action available, the processing chain continues. Otherwise, this implies the processing chain has ended.
\end{enumerate}

\subsection{TC Direct-Action Filter and clsact qdisc}
TC direct-action(da) can simply considered a method to use the filter as an action at the same time. In traditional TC filters, if we return some value like TC\_ACT\_SHOT(2), the value would be considered in the type of filter return value, i.e., the value would be seen as a classid. If we use the TC da, the return value would be considered as the TC action correctly. It avoids the need to create redundant TC actions and optimizes the performance of the TC subsystem because eBPF program-based filters have enough power to perform packet processing. 

TC \textit{clsact} is a pseudo qdisc, providing more programmability for the TC eBPF program. It does not perform any queuing, instead, it only contains filters and actions. It can be executed without acquiring the qdisc lock(lockless) since it works before the “real” qdisc, making primary packet processing tasks can be done earlier. Most importantly, it works on both the ingress and egress path, while the TC da only works on the ingress path, it lets TC da can now apply to the egress path qdisc too.

\section{Related Work}
\label{relate}
\subsection{Secure Modbus Protocol}
Numerous works have been dedicated to enhancing the security of the Modbus protocol. One notable solution involves implementing data integrity, authentication, non-repudiation and anti-replay features \cite{Design-and-Implementation-of-a-Secure-Modbus-Protocol}. In the work by Martins et al. \cite{Enhanced-Modbus-TCP-Security-Protocol}, the authors proposed the implementation of authentication and authorization functions. This was achieved through the integration of username and password-based access control, employing user-defined function codes. Xuan et al. \cite{Xuan_2019} proposed the Modbus-S protocol, which uses the AES algorithm to guarantee data confidentiality, a synchronization mechanism based on a hash algorithm for ensuring data uniqueness, and a digital signature algorithm to uphold data verifiability and integrity.

The Modbus organization has also released a new protocol specification named Modbus TCP Security \cite{Modbus-TCP-Security}. This protocol uses TLS to encapsulate traditional Modbus packets, providing both authentication and message-integrity protection. The protocol also introduces Role-Based Access Control (RBAC) information by leveraging the X.509v3 certificate extension.

\bibliographystyle{IEEEtran}
\bibliography{references}
\end{document}